\documentclass [prb,groupedadress,showpacs,twocolumn]{revtex4}
\usepackage{amsfonts, amsmath,graphicx,dcolumn,times}
\newcommand{\nn}{\nonumber\\}
\newcommand{\be}{\begin{eqnarray}}
\newcommand{\ee}{\end{eqnarray}}
\newcommand{\cht}{\tensor{\chi}}
\begin{document}
\author{C. Li}
\email{chun@physik.uni-kl.de}
\author{G. Lefkidis}
\author{W. H{\"u}bner}%
\affiliation{%
Department of Physics and Research Center OPTIMAS, Kaiserslautern University of Technology\\
PO Box 3049, 67653, Kaiserslautern, Germany }
\date{25.11.2008}
\title{Electronic theory of ultrafast spin dynamics}

\begin{abstract}
NiO is a good candidate for ultrafast magnetic switching because of its large spin density, antiferromagnetic order, and clearly separated intragap states. In order to detect and monitor the switching dynamics, we develop a systematic approach to study optical second harmonic generation (SHG) in NiO, both at the (001) surface and in the bulk. In our calculations NiO is modeled as a doubly embedded cluster. All intragap \emph{d}-states of the bulk and the (001) surface are obtained with highly-correlational quantum chemistry and propagated in time under the influence of a static magnetic field and a laser pulse. We find that demagnetization and switching can be best achieved in a subpicosecond regime with linearly rather than circularly polarized light. We also show the importance of including an external magnetic field in order to distinguish spin-up and spin-down states and the necessity of including magnetic-dipole transitions in order to realize the $\Lambda$-process in the centrosymmetric bulk. Having already shown the effects of phonons in the SHG for the bulk NiO within the frozen-phonon approximation, and following the same trail of thoughts, we discuss the role of phonons in a fully quantized picture as a symmetry-lowering mechanism in the switching scenario and investigate the electronic and lattice temperature effects.
\end{abstract}

\maketitle

\section{Introduction}

In recent decades there has been a continuous strain to
minimize as much as possible the time needed to record data on
magnetic materials. At the same time, the tremendous increase in storage density and read-write speed in magnetic storage media is reaching its physical limits. 
Ever since the light-induced demagnetization of ferromagnets was
discovered~\cite{bigot_demag}, several light-driven scenarios and
mechanisms have been proposed~\cite{koopmans,perakis}, and it has
already been demonstrated that exploiting the ultrafast
electron-photon interaction can lead to subpicosend
dynamics~\cite{abal}. Furthermore, it has been experimentally assessed that the
electronic excitations can non-thermally control spin
dynamics~\cite{naturekimel} and magnetic phases~\cite{naturemanfred}
in magnetic materials. Laser manipulation has been well understood in atomic and molecular systems,
which stimulates the demand for an implementation of these mechanisms in practical device applications. 
Thus, the investigation of electron and spin dynamics in the bulk, surfaces, and nanostructures of nonmagnetic as well as magnetic materials has attracted considerable interest in recent years.

Nonlinear optics is more selective than nonlocal linear optics, making it an ideal tool for investigating
antiferromagnets. NiO is a good candidate for such a scenario due to its
large spin density, antiferromagnetic (AF) order, and clearly separated
intragap states. Although bulk NiO possesses
a center of inversion, it generates a second harmonic
signal~\cite{manfred}. In the literature there are todate four
main explanations for its origin (i) spin-orbit coupling (SOC), 
(ii) a signal that mainly results from the surface, (iii) inclusion 
of higher order transitions, and (iv) lowering of the
crystallographic symmetry due to local distortions or phonons.
Recent works~\cite{lefkidis,KS_SOC} from our group showed that
explanations (i) and (iii), especially the latter, are the most 
probable candidates for the physics behind. The first two explanations 
are also demonstrated by a recent experiment~\cite{SHG2008}.

In this paper, we review our studies on electronic theory of ultrafast 
spin dynamics in recent years. The remainder of this paper is organized 
as follows. In Sec. 2, we introduce quantum chemistry methods used
in the calculations. In Sec. 3, we include SOC. In Sec. 4, 
second harmonic generation (SHG) is calculated. Symmetry analysis of the contributions
to the second order susceptibility tensors ($\cht^{(2\omega)}$) is performed. In Sec.~5, we show
the effects of phonons as a symmetry-lowering mechanism in the switching scenario. In Sec. 6,
we present our final results for realization of spin manipulation. 
Finally, in Sec. 7, we present our summary and outlook for future work.

\section{Quantum chemistry}

One can use either the real or the momentum space approach to model 
both the surface and the bulk of NiO.  Their major
difference is that only the first one is able to find the localized 
intragap \emph{d}-states of the Ni ion.  These
dispersionless states are experimentally confirmed for the surface
~\cite{fromme, 001_EELS} and for the bulk~\cite{optical_bulk}.
Although some previous calculations with extended local density
approximation (LDA++) within the Hubbard-I
approach~\cite{katsnelson} and within the $GW$
approximation~\cite{aryasetiawan,faleev,louie} give better results
for the gap than the real space approach, and a recent calculation combining 
LDA and dynamical mean-field theory provides an accurate band structure of NiO~\cite{VollhardtPRL}, 
they persistently miss the intragap states. Additionally, 
correlations play an important role in NiO, and the real space approach
is therefore more suitable to describe its complex electronic
structure. Last but not least we wish not only to calculate the
energy levels but the wavefunctions.

\subsection{Doubly embedded cluster}

\begin{figure}[t]
\includegraphics[width=8cm]{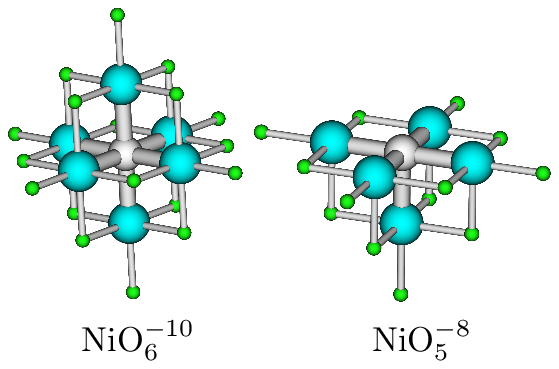}
 \caption{\label{structure} Clusters used for the
 bulk (left) and the surface (right). Large dark spheres represent
 oxygen atoms, the central grey sphere the nickel atom, and the
 small spheres the surrounding effective core potentials.
 }
\end{figure}

In order to model the system we use a doubly embedded
cluster~\cite{lefkidis,khompat}. For the surface a NiO$_5^{-8}$
cluster is used and for the bulk a NiO$_6^{-10}$ cluster. Both
clusters are first embedded in a shell of effective core
potentials (ECP), that account for the electrostatic properties of
the Ni ions in the immediate vicinity (Fig.~\ref{structure}).
The whole structure is then again embedded in a charge point field
(CPF) which describes the surrounding Madelung potential.
\subsection{Computational levels}
 We use several correlation levels in the calculations, mainly for
 comparison and checking reasons, such as configuration
 interaction (CI) with single excitations (CIS), single and double
 excitations (CISD), single excitations with energy
 contributions from higher excitations [CIS(D)]~\cite{cisd2},
 quadratic configuration interaction (QCI), with singles
 (QCIS), doubles (QCISD), and doubles with higher excitation energy
 contributions (QCISD-T).
 A method well suited to describe all \emph{d}-states
 is the multiconfigurational complete active space configuration
 interaction (MC-CAS or CASSCF), which can describe each state separately 
 and can include dynamical correlations as well~\cite{lefkidisPRB}.
 Table~\ref{table:energies} shows some results of NiO$_5^{-8}$ from 
 different correlation levels.

\begin{table*}
\caption{\label{table:energies}\emph{d}-state and gap energies (in eV) at various correlation levels of the NiO$_5^{-8}$ cluster~\cite{text}.}
\begin{tabular}{ccccccccccc}
\hline\hline
{NiO$_5^{-8}$}&{HF}&{$^3$CIS}&{$^3$CIS(D)}&{$^1$CIS}&{CCSD}&{QCISD(T)}&{CAS}&{exp.$^a$}&{NiO$_5^{-8}$}&{CAS}\\
\hline
 1$^3$B$_1$& 0.00 & 0.00 & 0.00 & 0.00 &  0.00 & 0.00 & 0.00 & 0.0&1$^1$B$_1$&  2.15\\
 1$^3$E    & 0.50 & 0.53 & 0.59 & 0.78 &   0.56 &  0.56 & 0.49& 0.6&2$^1$A$_1$&3.49\\
 1$^3$B$_2$& 0.93 & 2.00 & 1.74 &  \mbox{--}  &  1.14 &  1.17 & 0.92 &1.1&2$^1$E&3.93\\
 1$^3$A$_2$& 1.93 & 0.97 & 1.16 & 1.85 &  1.38 &  1.26 & 1.06 &1.3&3$^1$A$_1$&4.20\\
 2$^3$E    & 1.73 &  \mbox{--}  &  \mbox{--}  &  \mbox{--}  &  1.68 &  1.54 & 1.25 &\mbox{--}&2$^1$B$_1$&4.22\\
 1$^1$A$_1$& 2.79 &  \mbox{--}  &  \mbox{--}  & 2.61 &  1.47 & 1.26 & 1.82 &\mbox{--}&3$^1$E&4.23\\
 1$^1$E    & 5.12 &  \mbox{--}  &  \mbox{--}  & 2.82 &  3.10 & -4.75 & 2.51& 2.1&2$^1$B$_2$&4.35\\
 1$^1$B$_2$& 4.30 &  \mbox{--}  &  \mbox{--}  & 2.24 &  2.84 &  1.76 & 2.94&\mbox{--}&4$^1$A$_1$&8.29\\
  2$^3$A$_2$& 1.94 &  \mbox{--}  &  \mbox{--}  &  \mbox{--}  & 1.79 &  1.60 & 2.95 &\mbox{--}&\mbox{--} &\mbox{--}\\
 1$^1$A$_2$& 4.71 &  \mbox{--}  &  \mbox{--}  & 3.01 &  3.07 &  -2.69 & 3.15 &\mbox{--}&\mbox{--}&\mbox{--}\\
 3$^3$E    &  \mbox{--}  & 2.85 & 2.77 &  \mbox{--}    &  \mbox{\mbox{--}} &\mbox{--}  & 3.40 &\mbox{--}&\mbox{--}&\mbox{--}\\
  Gap&  7.19  &  7.64  &  5.57  &  10.76  & \mbox{--}&\mbox{--}  &  7.56 &\mbox{--}&\mbox{--}&\mbox{--}\\
\hline\hline
$^a$~Taken from Ref. 11.
\end{tabular}\\
\end{table*} 

\section{Spin-orbit coupling}

The SOC, characterized by the interaction between the electron spin
moment and the orbital angular momentum, results in an additional effect 
of splitting of the electron energy levels.
The high-quality linear and nonlinear (SHG) optical spectra of
NiO given by Fiebig \emph{et al.}~\cite{manfred} provide a good chance to
directly compare the experimental and theoretical details of
the fine structure.

Using the CIS method and including SOC, we concentrate on the 21
lowest triplet states originating from the crystal field and
spin-orbit splitting of the $^3F$ level of the Ni$^{2+}$ ion. 
Symmetry analysis shows that SOC does not split the singlet states.
It shifts, however, their energy due to the admixture of the triplet states. 
As a consequence some singlet-triplet transitions become allowed.

Figure 2 in reference~\cite{KS_SOC} shows the calculated fine structure
of bulk NiO. Including SOC leads to a 
slight lowering of the ground state energy.
The splitting of the second excited state ($\Delta$$E_{^3{\Gamma}_5^+}$=
$E_{{\Gamma}_2^+}$$-$$E_{{\Gamma}_3^+}$=71 meV) ($^3{\Gamma}_5^+$) is in good
agreement with the most accurate measurements (around 70 meV) of the optical absorption
and SHG spectra of NiO~\cite{manfred}.
In addition, our transition ${^3{\Gamma}_5^+}\rightarrow{^3{\Gamma}_2^+}$ at
about 0.9 eV agrees well with the SHG structure around
1.0 eV by Fiebig \emph{et al.}~\cite{manfred}.

We have also found that the splitting
of energies leads to additional electric dipole (ED) transitions, which corresponds
to the SHG response that can be observed when SOC is
taken into account.

\section{Second harmonic generation}

\subsection{Theory}

When we consider a monochromatic plane wave 
interacting with a material, it induces a response which leads to a
source term in the Maxwell equations. Expanding its inhomogeneous
solution yields terms of the polarization (${\bf P}_s$), the
magnetization (${\bf M}_s$), and the electric quadrupole (EQ)
polarization ($\tensor{Q}_s$), which are functions of the incident
field~\cite{sugano,shen}. Taking into account no more than one
second-order transition and considering the indistinguishability of the two
incident photons, we have
 \be
 \left(
 \begin{array}{c}
 \mathbf{P}_s\\
 \mathbf{M}_s\\
 \tensor{Q}_s\\
 \end{array}
 \right)\propto
\left(
 \begin{array}{lll}
\cht^{eee} & \cht^{eem} & \cht^{eeq} \\
\cht^{mee} & 0 & 0  \\
\cht^{qee} & 0 & 0 \\
\end{array}
 \right)\cdot
 \left(
 \begin{array}{cc}
 {\bf E}{\bf E}\\
 {\bf E}{\bf H}\\
 {\bf E}\tensor{Q}\\
 \end{array}
 \right)
 \ee
In the right side, the two last upper indices denote the
nature of the two absorptions, while the first one the nature of
the emissions [$e$ for ED, $m$ for magnetic dipole (MD), and $q$
for EQ induced transitions], and the lower indices
in the subsequent formulas represent the directions of the
corresponding multipoles. To calculate $\cht^{(2\omega)}$, 
adopting the Coulomb gauge, neglecting the $\mathbf{A}^2$ term
(this term leads to a tensor completely analogous to the 
nonlinear optical conductivity tensor $\tensor{\Sigma}^C$ 
discussed by Andersen \emph{et al.}~\cite{TA-nonlinear}) 
and setting
${\bf n}$ and ${\bf a}$ as the unit vectors
 in the propagation and the polarization direction
 of the light beam respectively,
 we get a perturbation term, a Taylor expansion of which
 keeping only the first two
 terms in $i\frac{\omega}c{\bf n}\cdot{\bf r}$ yields for the interaction elements:
  \be
V'_{ij}&\simeq&-\frac{e\omega_{ij}{\cal
E}}{\omega}d_{ij}^{\bf{a}}-\frac{e{\cal
B}}{2m_ec}L_{ij}^{\bf{a}\times\bf{n}}
-\frac{e\omega_{ij}\nabla\cdot{\cal
E}}{2\omega}Q_{ij}^{\bf{a},\bf{n}}
 \ee
where ${\cal E}$ and ${\cal B}$ are magnitudes of the electric and
magnetic fields, respectively, $\omega$ is the frequency of the
incident light, $d_{ij}^{\bf{a}}$,
 $L_{ij}^{\bf{a}\times\bf{n}}$ and $Q_{ij}^{\bf{a},\bf{n}}$ are the
components of the ED, MD, and EQ transitions along the directions indicated by the
superscripts. The microscopic formulas for $\cht^{(2\omega)}$, 
derived from the Liouville equation of motion, by
applying second order perturbation theory~\cite{khompat,shen,tensorformula},
has already been given~\cite{lefkidis,lefkidisPRB}
 \be\label{SHG}
 \chi^{ijk}_{xyz}&\propto&\sum_{\alpha\beta\gamma}
\Bigg[
V^{i,x}_{\gamma\alpha}\overline{V^{j,y}_{\alpha\beta}V^{k,z}_{\beta\gamma}}\times\nn
 &&\frac{\frac{f(E_{\gamma})-f(E_{\beta})}{E_{\gamma}-E_{\beta}-\hbar\omega+i\hbar\Gamma}-
\frac{f(E_{\beta})-f(E_{\alpha})}{E_{\beta}-E_{\alpha}-\hbar\omega+i\hbar\Gamma}}
 {E_{\gamma}-E_{\alpha}-2\hbar\omega+2i\hbar\Gamma}\Bigg],
 \ee
where the indices \emph{i}, \emph{j}, \emph{k} denote the nature of the
 transition (i.e., \emph{e} for ED, \emph{m} for MD, and \emph{q} for EQ transitions) 
between states $\alpha$ and $\beta$, and $f(E)$ is the population
distribution function.
The overbar means symmetrization with respect to the two incident photons.

\subsection{Symmetry analysis}

Our cluster is in the ferromagnetic phase due to the model involving only one Ni atom.
The real space approach is not able to describe the translational invariance of the real
crystal lattice and the different domains, which give different
second harmonic signals~\cite{aalborg,daehnPRL}.

The NiO$_5^{-8}$ cluster has a C$_{4\text{v}}$ symmetry with no inversion center, and
 therefore all rank three tensors are allowed, and their
 symmetrized forms are
 \be
 \cht^{eee}_{C_{4\text{v}}}=\left(
 \begin{array}{ccc|ccc}
 0&0&0&0&\chi_{xxz}^{eee}&0\\
0&0&0&\chi_{xxz}^{eee}&0&0\\
\chi_{zxx}^{eee}&\chi_{zxx}^{eee}&\chi_{zzz}^{eee}&0&0&0
 \end{array}
 \right),
 \ee
 \be
\cht^{mee}_{C_{4\text{v}}}=\left(
 \begin{array}{ccc|ccc}
 0&0&0&\chi^{mee}_{xyz}&0&0\\
0&0&0&0&-\chi^{mee}_{xyz}&0\\
0&0&0&0&0&0
 \end{array}
 \right),
 \ee
 \begin{widetext}
and
 \be
\cht^{eem}_{C_{4\text{v}}}=\left(
 \begin{array}{ccc|ccc|ccc}
 0&0&0 & \chi^{eem}_{xyz}&0&0&  \chi^{eem}_{xzy}&0&0\\
 0&0&0 & 0&-\chi^{eem}_{xzy}&0  & 0&-\chi^{eem}_{xyz}&0\\
 0&0&0& 0&0&\chi^{eem}_{zxy} &0&0&-\chi^{eem}_{zxy}
 \end{array}
 \right)
 \ee
 \end{widetext}
with three, one, and three independent matrix elements, respectively. 
Symmetrizing results in the vanishing of the $\chi^{eem}_{zxy}$ element, in the
O$_\text{h}$ point group (bulk) we further have
$\chi^{eem}_{xyz}=-\chi^{eem}_{xzy}$ and thus the whole tensor
disappears. 

The tensors are of rank four when involving EQ transitions.
Symmetrizing with respect to the last two indices can simplify the $\cht^{qee}$ tensor  
both for the bulk and for the surface, $\chi^{qee}_{(ij)kl}$ has
nine nonvanishing contributions. For the group O$_\text{h}$ we are
treating with a T$_4$ tensor~\cite{birss} with three independent
elements, the symmetrized form is
\begin{widetext}
\be
 \chi^{qee}_{O_\text{h}}&=&\left(
 \begin{array}{ccc|ccc}
 \chi^{qee}_{(x^2-y^2)xx}&\chi^{qee}_{(x^2-y^2)xx}&0&
 0&0&0\\
 \chi^{qee}_{(z^2)xx}&\chi^{qee}_{(z^2)xx}&\chi^{qee}_{(x^2-y^2)xx}+\chi^{qee}_{(z^2)xx}&
 0&0&0\\
 \hline
 0&0&0&
 \chi^{qee}_{(xy)xy}&0&0\\
 0&0&0&
 0&\chi^{qee}_{(xy)xy}&\\
 0&0&0&
 0&0&\chi^{qee}_{(xy)xy}\\
\end{array}
 \right).
 \ee
\end{widetext}

Furthermore we have also analyzed, using group theory, the contributions to the $\cht^{(2\omega)}$
arising from magnetic domains rotated with respect to each other~\cite{aalborg}. 
In addition, a similar symmetry analysis
has been performed for ferromagnetic systems to obtain the nonlinear susceptibility 
tensor by Andersen \emph{et al.} where even and odd contributions are attributed to
spin-dependent transitions~\cite{AndersenSHG}.

\subsection{Results}

\begin{figure}[t]
\includegraphics[width=8cm]{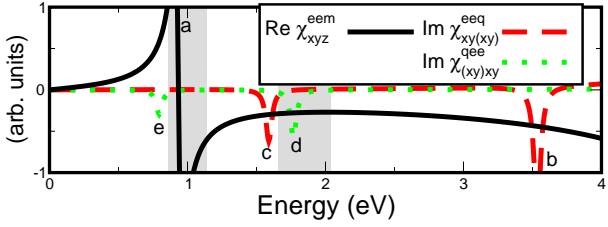}
\caption{\label{fig:bulk} Tensor elements of the
bulk of NiO. Peak $a$ corresponds to transition $1^3\text{A}_{2g}
\leftrightarrow 1^3\text{T}_{2g}$ with frequency $\omega$, peak $b$
to $1^3\text{A}_{2g} \leftrightarrow 2^3\text{T}_{1g}$ with
frequency $\omega$, peak $c$ to $1^3\text{A}_{2g} \leftrightarrow
1^3\text{T}_{1g}$ with frequency $\omega$, peak $d$ to
$1^3\text{A}_{2g} \leftrightarrow 2^3\text{T}_{1g}$ with frequency
$\omega$ and peak $e$ to $1^3\text{A}_{2g} \leftrightarrow
1^3\text{T}_{1g}$ with frequency $2\omega$.}
\end{figure}

Altogether there are three types of independent tensor elements, 
namely $\chi^{eem}_{xyz}$, $\chi^{eeq}_{xy(xy)}$, 
and $\chi^{qee}_{(xy)xy}$~\cite{birss}, in bulk NiO. The
most distinct peak at 0.94 eV (Fig.~\ref{fig:bulk}) due to the MD 
transitions perfectly coincides with the
experimental results~\cite{manfred}. Peak \emph{d}, which arises from an
EQ transition, also agrees very well. 
Further splitting of these peaks can easily be attributed to
SOC and/or phonons.
\begin{figure}
\includegraphics[width=8cm]{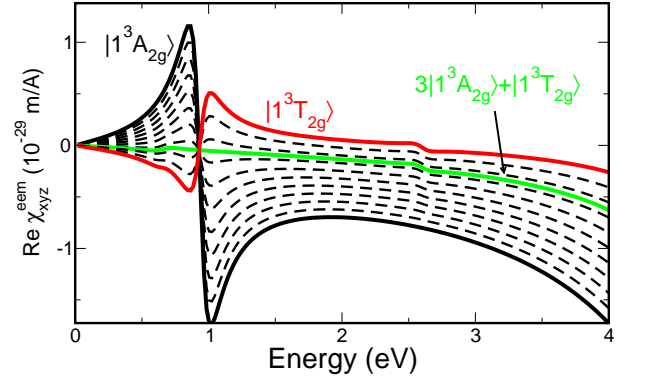}
\caption{\label{fig:thermal} Energetically balanced
repopulation of the NiO$_6^{-10}$ cluster and its effect on the
$\chi^{eem}_{xyz}$ tensor element. The outer solid lines (black and red) represent the
tensor element for fully populated ground and first excited state, 
the inner solid line (green) the suppressed peak when the population ratio 
equals 3:1.}
\end{figure} 

\begin{figure}
\includegraphics[width=8cm]{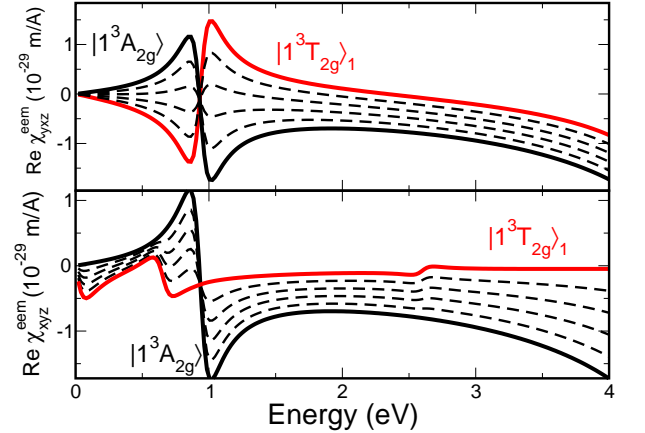}
\caption{\label{fig:nonthermal} Energetically
nonbalanced repopulation of the NiO$_6^{-10}$ cluster and its
effect on the $\chi^{eem}_{xyz}$ and $\chi^{eem}_{yxz}$ tensor
elements.}
\end{figure}

Considering certain experimental conditions,
we repopulate the states in two ways~\cite{lefkidisPRB}, that is, 
an energetically balanced way which consists of an equal distribution 
of electrons in all sublevels with a given energy, and an energetically 
nonbalanced way where anisotropy is included at the same time, as 
demonstrated in Figures~\ref{fig:thermal} and~\ref{fig:nonthermal}. 
The first case could happen when a system is prepared with an unpolarized 
pump laser pulse, and the results show that higher kinetic temperatures would
result in excitations to different states. In the second way, which could 
happen with a polarized pump beam, $\chi^{eem}_{yxz}$ shows the same behavior as in
the thermal distribution, while $\chi^{eem}_{xyz}$ exhibits
additional symmetry violating features.
\section{Phonons}

\subsection{Frozen-phonon approximation}

Up to now, we supposed a non-distorted 
lattice, which of course is not true in the real system. 
However, the system deviates from the full O$_\text{h}$ symmetry through
the static lattice distorsion along the $\langle 111\rangle$ axis due 
to spin ordering or dynamic distortions like phonons. 
Here we consider the latter case, which takes place both in the bulk 
and the surface. We treat the electronic movements quantum mechanically 
within the Born-Oppenheimer approximation while the 
motion of the nuclei is considered to happen on a much slower pace, 
giving time to the electrons to adiabatically follow them.

\begin{figure}[t]
\setlength{\unitlength}{1cm}
\begin{picture}(8,7)
\put(0,0){\includegraphics[width=8cm]{figure5.eps}}
\put(3.7,4.1){\includegraphics[width=1.2cm]{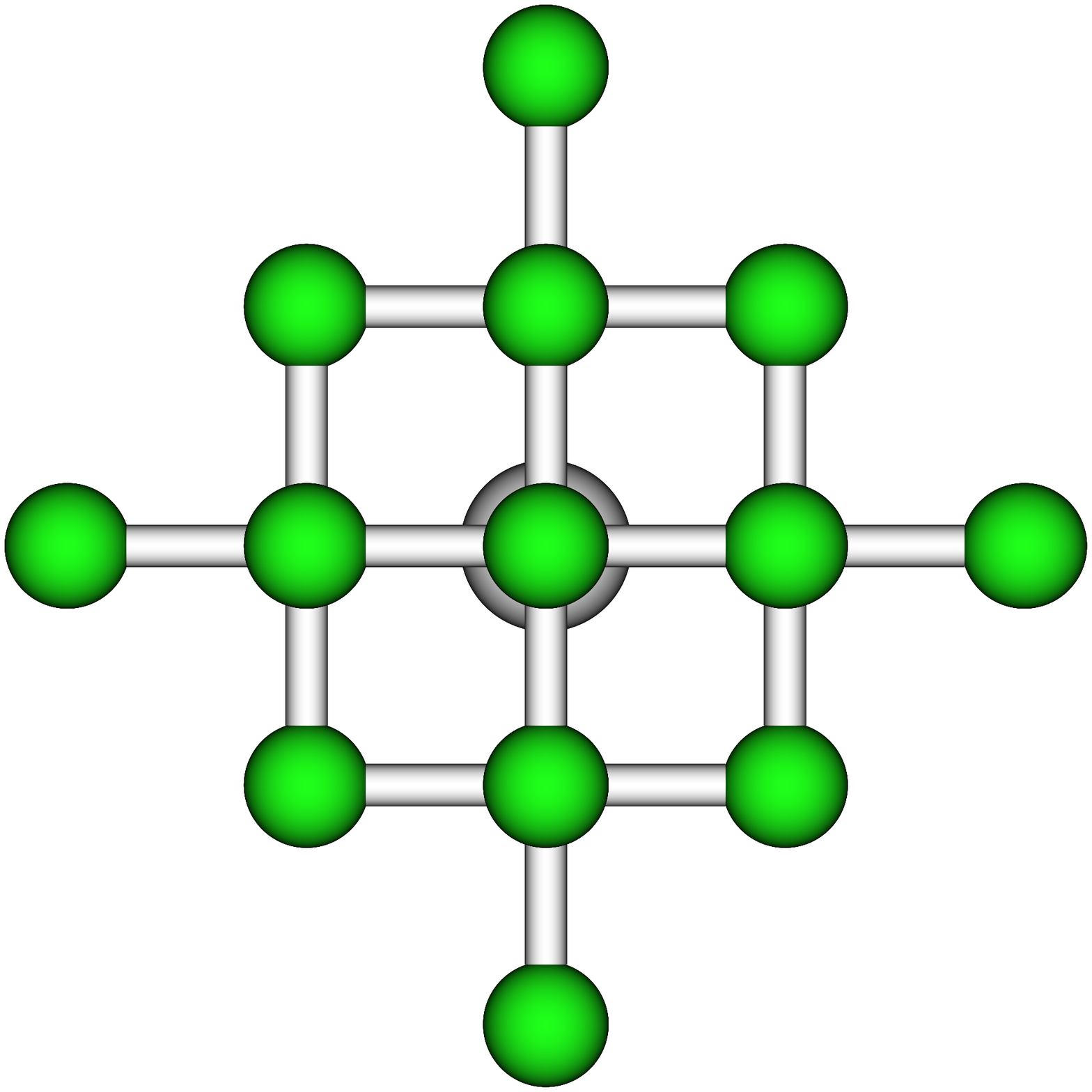}}
\put(5.85,4.1){\includegraphics[width=1.2cm]{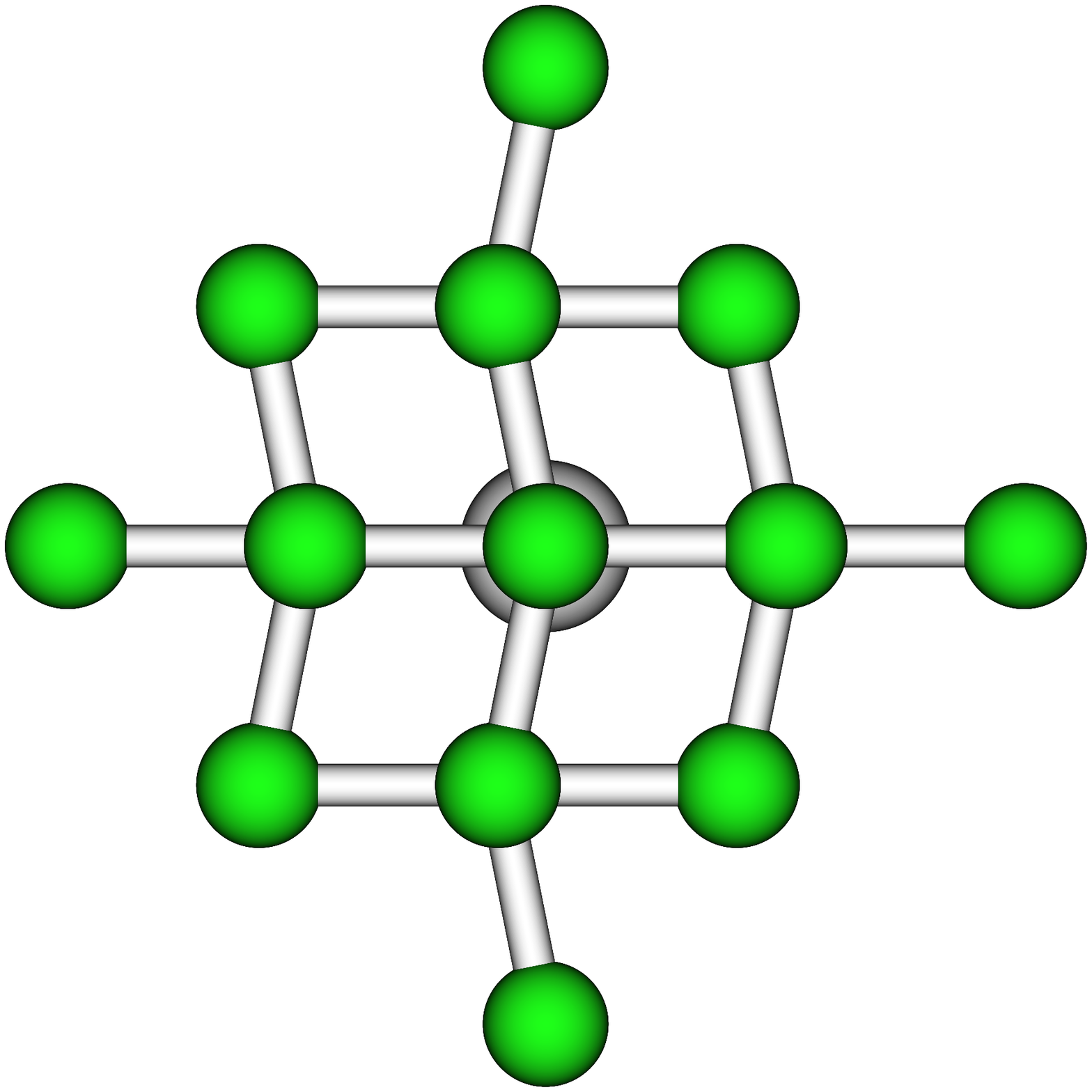}}
\put(1.55,4.1){\includegraphics[width=1.2cm]{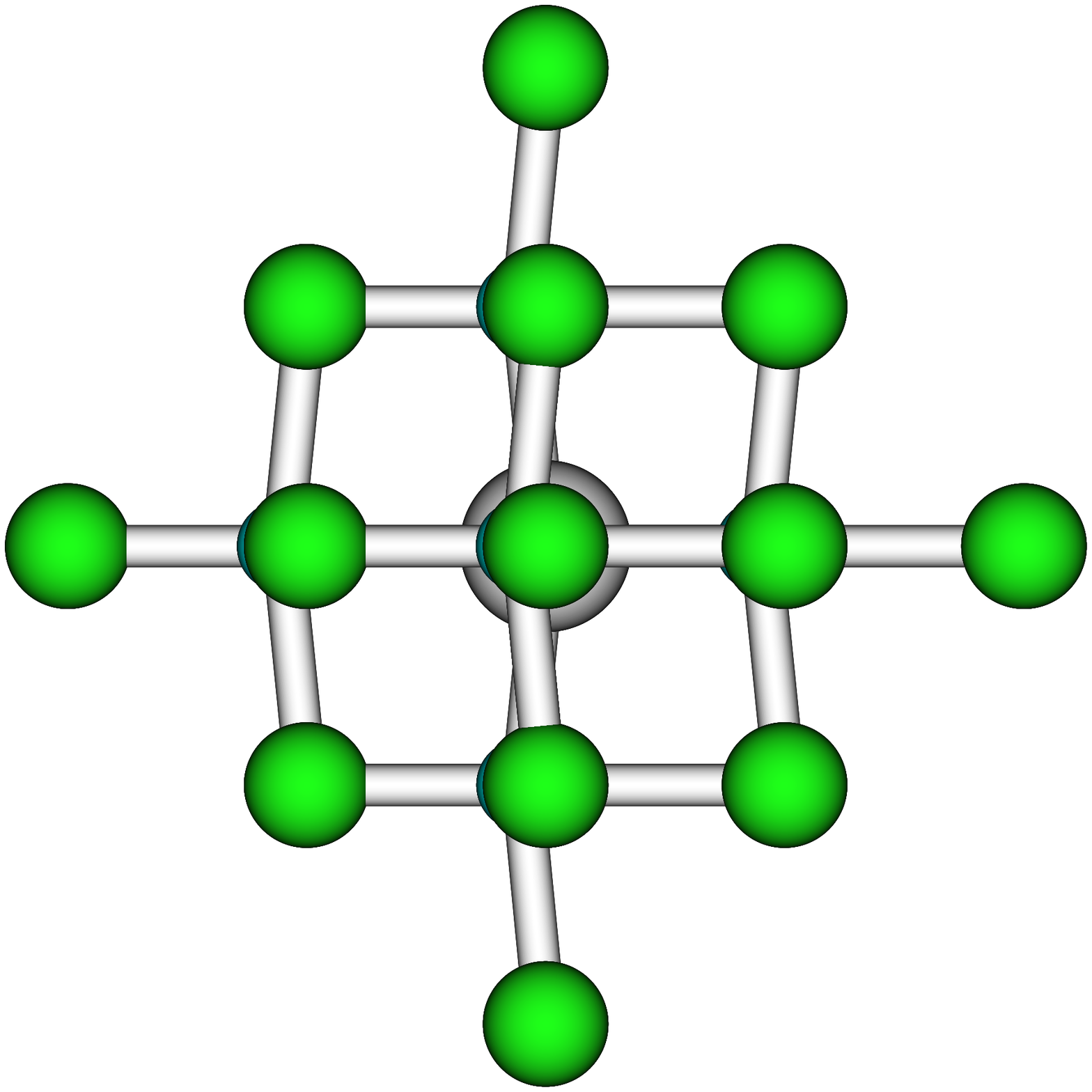}}
\end{picture}
\caption{\label{phonon-levels} In the middle panel
are the calculated levels for O$_\text{h}$ symmetry, on the left
for C$_{4\text{v}}$ and on the right for C$_{2\text{v}}$. The
insets show the respective clusters (with the effective charge
points).}
\end{figure}

The optical phonons lower the local symmetry of the NiO$_6^{-10}$ in our rocksalt structure,
even at the $\Gamma$ point.
There are a few cases, however, where the local geometry on any Ni site is the same 
throughout the crystal, the level splittings of which are shown in Fig.~\ref{phonon-levels}. 

\begin{figure}[t]
\includegraphics[width=8cm]{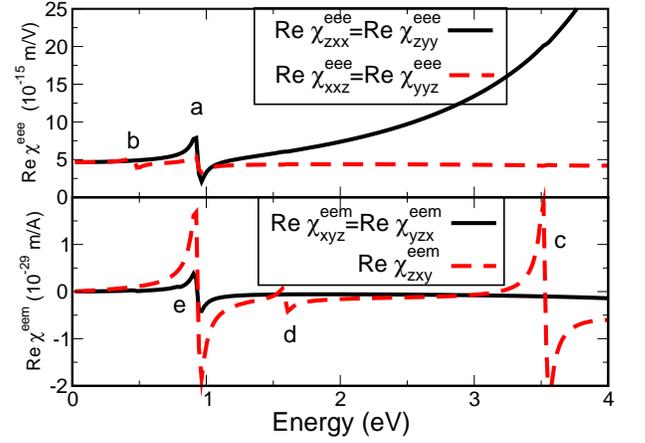}
\caption{\label{fig:x-C4v} $\cht^{eee}$ and
$\cht^{eem}$ tensors for the NiO$_6^{-10}$ in the case of the
$\Gamma$ point of one optical phonon along the $z$ direction
(C$_{4\text{v}}$ symmetry). Note that peaks \emph{a} and \emph{e} 
are shifted with respect to each other by 5 meV, a splitting which 
is visible only if zoomed in.}
\end{figure}

Now SHG becomes possible even within the ED
approximation since the inversion symmetry is broken.
Fig.~\ref{fig:x-C4v} depicts the $\cht^{eee}$ and the $\cht^{eem}$
tensor elements in the NiO$_6^{-10}$ cluster for a phonon
coordinate of 3\% dislocation, at the $\Gamma$ point, when the
symmetry on every Ni site is C$_{4\text{v}}$. Here, the energy splitting 
of the peak near 1 eV of the
$\cht^{eem}$ (about 5 meV) is very close to the experimental
splitting of the first intense SHG peak observed by Fiebig
\emph{et~al.}~\cite{manfred}.

\begin{figure}[t]
\includegraphics[width=5cm]{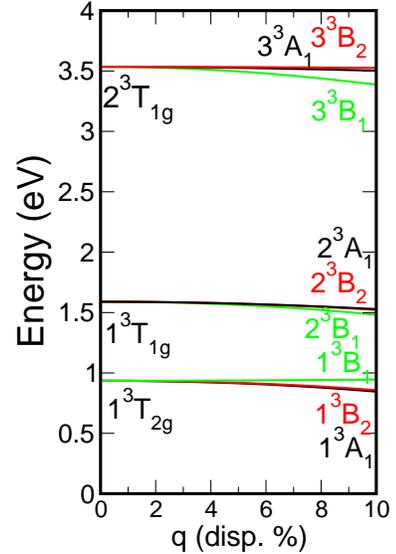}
\caption{\label{fig:levels} Splitting of the levels
vs. displacement in the case of the transversal plane wave (see
text) along the $z$ direction (C$_{2v}$ symmetry). The splitting is
an even function of \textbf{q}.}
\end{figure}

A lower symmetry will lead to an even richer structure~\cite{lefkidisPRB}. 
Note that generally the triply degenerate
states get split into three levels, but two of which are always
almost degenerate in energy, and thus the peaks get split into two
peaks and not three, as shown in Fig.~\ref{fig:levels}.
\subsection{Time evolution}

\begin{figure}[t]
\includegraphics[width=8cm]{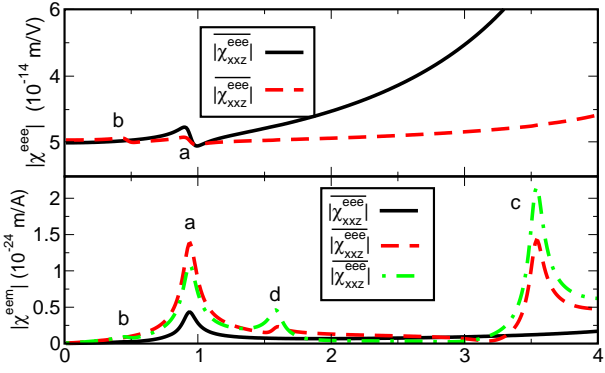}
\caption{\label{fig:chi-averaged} Time averaged
$|\cht^{eee}|$ and $|\cht^{eem}|$ tensors for the NiO$_6^{-10}$ in
the case of the $\Gamma$ point of one optical phonon along the $z$
direction (C$_{4\text{v}}$ symmetry) with an lattice displacement of
amplitude of 3\%.}
\end{figure}

\begin{figure}[t]
\includegraphics[width=8cm]{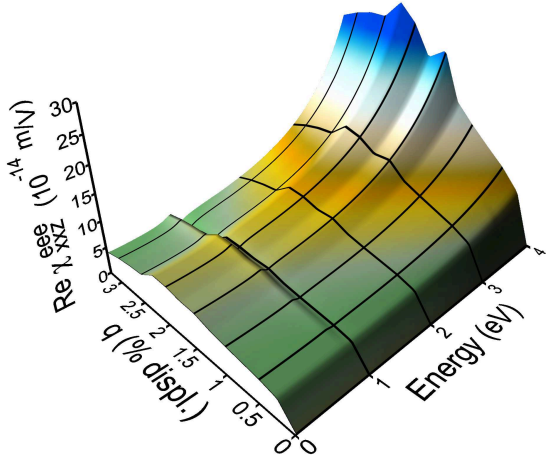}
\caption{\label{fig:time-xeee} $\chi^{eee}_{xxz}$
vs. photon energy and vs. \textbf{q}. A peak starts appearing
after ${\bf q}=0.25\%$ at energy $\sim 0.95$ eV.}
\end{figure}

\begin{figure}[t]
\includegraphics[width=8cm]{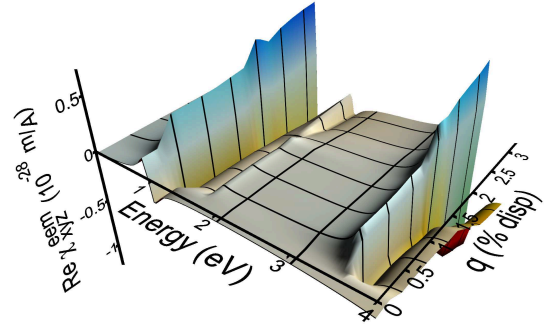}
\caption{\label{fig:time-xeem} $\chi^{eem}_{xyz}$
vs. photon energy and vs. \textbf{q}. Two new peaks appear after
${\bf q}=0.25\%$, one at energy $\sim 1.590$ eV which changes
its phase at ${\bf q}=2.0\%$, and one at energy $\sim 3.65$ eV.}
\end{figure}

The calculations show that the relation between 
$\cht^{(2\omega)}({\bf q})$ and \textbf{q} is not linear
(Fig.~\ref{fig:levels}). Nonetheless, since in our approximation we 
consider SHG to be instantaneous we calculate the time average 
over a whole period of $(\cht^{(2\omega)})^2$ as shown in
Fig.~\ref{fig:chi-averaged}. We consider a simple harmonic model
(uncoupled phonons), and take snapshots for amplitude of up to 3\%
lattice displacement with a step of 0.25\%. It is found that 
the peaks do not get broadened but split, 
and the intensity of the peaks has an
almost quadratic relationship to the amplitude of the
displacement.
The time resolved results shown in Fig.~\ref{fig:time-xeee}
and Fig.~\ref{fig:time-xeem} exhibit the relationship between the
tensors elements ($\chi^{eee}_{xxz}$ and $\chi^{eem}_{xyz}$) 
and the phononic coordinate~\textbf{q}). Both the behaviors are clearly
nonlinear, and one can distinguish three regimes~\cite{lefkidisPRB}. 
A similar behavior can be observed for the $\chi^{mee}_{xyz}$ tensor element.

\section{Spin manipulation}

\subsection{Simple models for laser-induced dynamics}

We start our investigations of the spin manipulation with
the simplest two-level and three-level systems~\cite{abalPRB}.
To obtain a spin mixed state, two excited 
states of definite (and opposite) spin are needed, which 
are mixed through SOC. Thus, the minimum scenario for 
optically induced spin dynamics is a four-level system:
two almost degenerate pure spin states at low energy (ground states) and 
two also almost degenerate states at high energy which can 
be mixed by SOC~\cite{abalJPCM}.

\subsection{Ultrafast magnetic switching in NiO}

\begin{figure}[t]
\includegraphics[width=8cm]{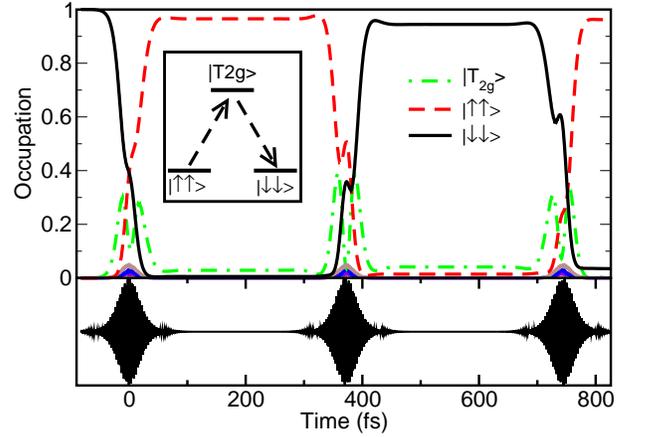}
\caption{\label{fig:many_pulses} Top: population
evolution of the lowest lying levels in bulk of NiO. Bottom: Laser profile.
}
\end{figure}

Based on the above suggested four-level system, we present an 
ultrafast magneto-optical switching mechanism in NiO. 
Fig.~\ref{fig:many_pulses} shows the population evolution
of the lowest levels in bulk NiO and the corresponding laser profile.
In order to achieve switching we have to choose a suitable
mixed-spin excited state. We find that the maximum efficiency occurs 
when the matrix elements from both ground states have almost the
same absolute values. The phase evolution of the different states governs
the direction of the population transfer during the pulse. That is why
Fig.~\ref{fig:many_pulses} exhibits a double-peak structure for the intermediate excited states.

\begin{figure}[t]
\includegraphics[width=8cm]{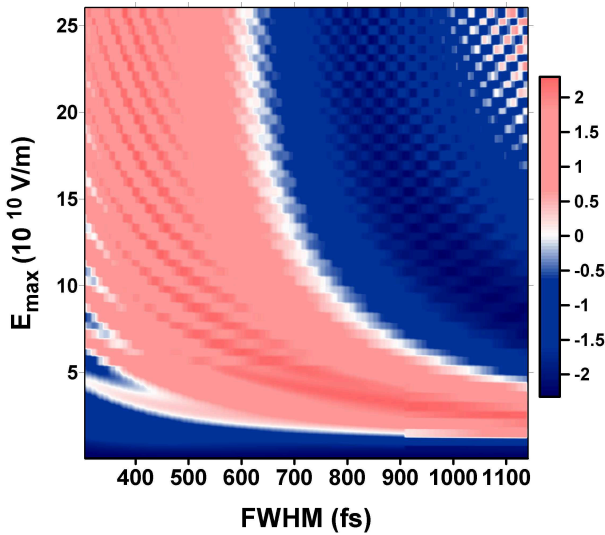}
\caption{\label{fig:hyperbolic} Contour plot of the final magnetic state of the
NiO$_5^{-8}$ cluster after the application of a linearly polarized
$sech^2$-shaped laser pulse and an external static magnetic field
$B_{\text{ext}}=B_z=12.5$ A/m vs.~duration and intensity of the
pulse.}
\end{figure}

Fig.~\ref{fig:hyperbolic} shows the dependence of the spin
magnetic moment of the NiO(001) surface with respect to the duration 
and intensity of the pulse, after the pulse is switched off and the 
system has reached a static state. 
The static B-field is
perpendicular to the surface and the laser pulse is
linearly polarized at a direction parallel to the surface.
Contrary to previous works~\cite{abal,abalPRB}, taking into advantage
of the breaking of the time reversal sysmmetry due to the B-field, 
we find optimal switching conditions for linearly
polarized light, more specifically with propagation direction
parallel to the static B-field---this is consistent with the
experimental findings of Koopmans \emph{et al.}~\cite{koopmanscondmat} 
in an experimental setup similar to ours. Our calculations indicate that the
use of circularly polarized light can selectively activate one
channel of the $\Lambda$-process (see
inset in Fig.~\ref{fig:many_pulses}), only allowing either excitation
or de-excitation, while a superposition of the two polarizations can selectively control the
percentage of the population transfer~\cite{lefkidisswitch}.
Similar to our results, Stanciu \emph{et al.} observed subpicosecond
magnetization reversal in a ferrimagnetic system, however, 
with a circularly polarized laser pulse without 
magnetic field~\cite{KimelPRL}.

Note that breaking of the time-reversal symmetry leads to a differentiation of the roles of 
$\langle L^2 \rangle$ and $\langle L \rangle$. While the former describes 
crystalline anisotropy and survives unharmed the existence of the external B-field, 
the latter one lifts spatial degeneracies and gives rise to dynamic effects of magnetic 
origin (e.g. Kerr and Faraday effects). Thus linear and quadratic coupling to the 
(anti)-ferromagnetic order parameters become detectable in optics.

\subsection{Quantized phonons}

\begin{figure}
\includegraphics[width=8cm]{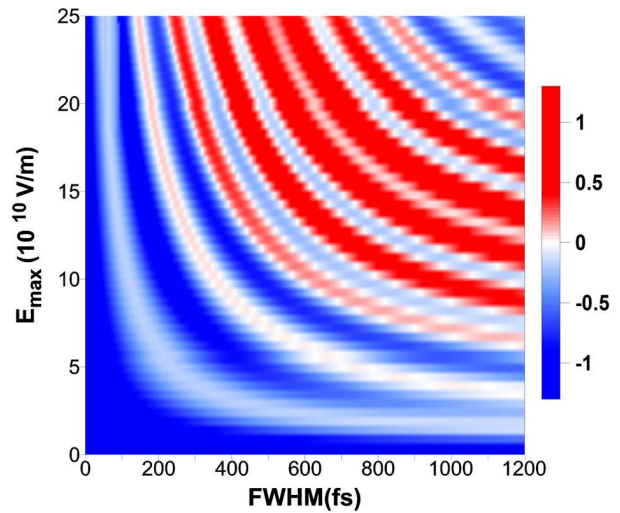}
\caption{\label{fig:quantized}Contour plot of the magnetic state of the
NiO$_6^{-10}$ cluster in the presence of one $\Gamma$-point
optical phonon after the application of a linearly polarized
$sech^2$-shaped laser pulse and an external static magnetic field
$B_{\text{ext}}=B_z=12.5$ A/m vs.~duration and intensity of the
pulse.}
\end{figure}

In order to consider phonons in a fully quantized picture we first calculate 
the normal modes of our cluster by diagonalizing the dynamical matrix. 
Then we quantize the normal modes and subsequently calculate the 
phonon-electron interaction by fully quantum-mechanically considering 
$\frac{\partial\langle\hat{H}\rangle}{\partial q}$, where $q$ is the phononic coordinate as seen in Figures~\ref{fig:time-xeee} and~\ref{fig:time-xeem}~\cite{QuantizedPhonon}. 
Then we include them together with the SOC and the external
static magnetic field in one diagonalization step and therefore to
eliminate the problem of applying twice a perturbative method on
the wavefunction. Fig.~\ref{fig:quantized} shows the magnetic state of the
NiO$_6^{-10}$ cluster in the presence of one $\Gamma$-point
optical phonon after the application of a linearly polarized 
laser pulse and an external static magnetic field vs.~duration and 
intensity of the pulse. The existence of optical phonons moves the switching
parameters towards higher intensities and longer times, while decreasing 
the tolerance of the system for achieving switching.

Temperature analysis indicates that the elevated lattice temperature 
makes the effect of switching more efficient, while electronic temperature (loosely
defined) does not significantly alter the process, since
the first one significantly changes symmetry, while the second one does not~\cite{QuantizedPhonon}. 

\section{Summary and outlook}

In summary, we have presented a quantum chemistry method for calculating different 
contributions to SHG both for the bulk and for the (001) surface of NiO, 
without involving any empirical parameter. We consider not only the ED approximation but 
MD and EQ induced transitions as well, and we also calculate 
the phononic effects within the frozen-phonon approximation. 
It is shown that all of the following, the SOC, the surface, the nonlocal contributions of the laser field,
and the transient lattice distortions, can give rise to peaks in the centrosymmetric bulk NiO. 
The results also indicate the possibility to
disentangle the surface (001) from the bulk signal due to
different geometrical and frequency dependencies, which has been
confirmed in a recent experiment~\cite{SHG2008}.

We present a fully \emph{ab initio} ultrafast magnetooptical switching 
mechanism in NiO and show that controlled switching is possible by 
including SOC in order to take advantage of a $\Lambda$-process. 
The role of phonons is further discussed in a fully quantized picture 
as a symmetry-lowering mechanism in the switching scenario.

The role of symmetry analysis is definitely significant through the 
whole paper. It has been used to classify the electronic states and judge
the allowed dipole transitions in SOC, simplify the tensor elements of the high
order transitions in SHG, and help us recognize the allowed transitions 
in the realization of ultrafast switching. 

Up to now, we only have considered one magnetic center (one Ni atom) 
in our investigations. Although the present single-site model can mostly 
account for AF ordering in NiO (both bulk and surface) 
because of the high spin density on the Ni sites, it is still necessary to
involve more magnetic centers to study more subtle systems for possible 
applications such as magnetic logic units. 

Note that since our calculations were performed on a one-magnetic-center cluster we
therefore did not need to consider gauge including atomic orbitals (GIAOs).
However, clusters containing more than one active magnetic center should (perhaps)
include the phase factor $e^{i\mathbf{A}\cdot{\mathbf{r}}}$ in the atomic orbitals.

Although spin is not a good quantum number after the inclusion 
of SOC the breaking of the time-reversal symmetry does not inhibit the 
calculation of its expectation value. However, it is necessary to account 
for it in a full spinor basis (two-component) even though rotational 
invariance is violated in the cluster. For systems where the spin density 
is localized on one magnetic center only (i.e. NiO) one can consider this 
center as the origin of the spatial symmetry operations, but for multi-centered 
clusters the necessity of fully accounting for the spinors becomes imminent.
 
\section*{Acknowledgements}

We would like to acknowledge support from the MINAS Landesschwerpunkt, and
Priority Programme 1133 of the German Research Foundation.

\bibliography{papers}

\begin{thebibliography}{36}
\expandafter\ifx\csname natexlab\endcsname\relax\def\natexlab#1{#1}\fi
\expandafter\ifx\csname bibnamefont\endcsname\relax
  \def\bibnamefont#1{#1}\fi
\expandafter\ifx\csname bibfnamefont\endcsname\relax
  \def\bibfnamefont#1{#1}\fi
\expandafter\ifx\csname citenamefont\endcsname\relax
  \def\citenamefont#1{#1}\fi
\expandafter\ifx\csname url\endcsname\relax
  \def\url#1{\texttt{#1}}\fi
\expandafter\ifx\csname urlprefix\endcsname\relax\def\urlprefix{URL }\fi
\providecommand{\bibinfo}[2]{#2}
\providecommand{\eprint}[2][]{\url{#2}}

\bibitem[{\citenamefont{Beaurepaire et~al.}(1996)\citenamefont{Beaurepaire,
  Merle, Daunois, and Bigot}}]{bigot_demag}
\bibinfo{author}{\bibfnamefont{E.}~\bibnamefont{Beaurepaire}},
  \bibinfo{author}{\bibfnamefont{J.-C.} \bibnamefont{Merle}},
  \bibinfo{author}{\bibfnamefont{A.}~\bibnamefont{Daunois}}, \bibnamefont{and}
  \bibinfo{author}{\bibfnamefont{J.-Y.} \bibnamefont{Bigot}},
  \bibinfo{journal}{Phys. Rev. Lett.} \textbf{\bibinfo{volume}{76}},
  \bibinfo{pages}{4250} (\bibinfo{year}{1996}).

\bibitem[{\citenamefont{Koopmans et~al.}(2005)\citenamefont{Koopmans, Ruigrok,
  Longa, and de~Jonge}}]{koopmans}
\bibinfo{author}{\bibfnamefont{B.}~\bibnamefont{Koopmans}},
  \bibinfo{author}{\bibfnamefont{J.~J.~M.} \bibnamefont{Ruigrok}},
  \bibinfo{author}{\bibfnamefont{F.~D.} \bibnamefont{Longa}}, \bibnamefont{and}
  \bibinfo{author}{\bibfnamefont{W.~J.~M.} \bibnamefont{de~Jonge}},
  \bibinfo{journal}{Phys. Rev. Lett.} \textbf{\bibinfo{volume}{95}},
  \bibinfo{pages}{267207} (\bibinfo{year}{2005}).

\bibitem[{\citenamefont{Chovan et~al.}(2006)\citenamefont{Chovan, Kavousanaki,
  and Perakis}}]{perakis}
\bibinfo{author}{\bibfnamefont{J.}~\bibnamefont{Chovan}},
  \bibinfo{author}{\bibfnamefont{E.~G.} \bibnamefont{Kavousanaki}},
  \bibnamefont{and} \bibinfo{author}{\bibfnamefont{I.~E.}
  \bibnamefont{Perakis}}, \bibinfo{journal}{Phys. Rev. Lett.}
  \textbf{\bibinfo{volume}{96}}, \bibinfo{pages}{057402}
  (\bibinfo{year}{2006}).

\bibitem[{\citenamefont{G{\'o}mez-Abal
  et~al.}(2004)\citenamefont{G{\'o}mez-Abal, Ney, Satitkovitchai, and
  H{\"u}bner}}]{abal}
\bibinfo{author}{\bibfnamefont{R.}~\bibnamefont{G{\'o}mez-Abal}},
  \bibinfo{author}{\bibfnamefont{O.}~\bibnamefont{Ney}},
  \bibinfo{author}{\bibfnamefont{K.}~\bibnamefont{Satitkovitchai}},
  \bibnamefont{and}
  \bibinfo{author}{\bibfnamefont{W.}~\bibnamefont{H{\"u}bner}},
  \bibinfo{journal}{Phys. Rev. Lett.} \textbf{\bibinfo{volume}{92}},
  \bibinfo{pages}{227402} (\bibinfo{year}{2004}).

\bibitem[{\citenamefont{Kimel et~al.}(2005)\citenamefont{Kimel, Kirilyuk,
  Usachev, Pisarev, Balashov, and Rasing}}]{naturekimel}
\bibinfo{author}{\bibfnamefont{A.~V.} \bibnamefont{Kimel}},
  \bibinfo{author}{\bibfnamefont{A.}~\bibnamefont{Kirilyuk}},
  \bibinfo{author}{\bibfnamefont{P.~A.} \bibnamefont{Usachev}},
  \bibinfo{author}{\bibfnamefont{R.~V.} \bibnamefont{Pisarev}},
  \bibinfo{author}{\bibfnamefont{A.~M.} \bibnamefont{Balashov}},
  \bibnamefont{and} \bibinfo{author}{\bibfnamefont{T.}~\bibnamefont{Rasing}},
  \bibinfo{journal}{Nature} \textbf{\bibinfo{volume}{435}},
  \bibinfo{pages}{6558} (\bibinfo{year}{2005}).

\bibitem[{\citenamefont{Lottermoser et~al.}(2004)\citenamefont{Lottermoser,
  Lonkai, Amann, Hohlwein, Ihringer, and Fiebig}}]{naturemanfred}
\bibinfo{author}{\bibfnamefont{T.}~\bibnamefont{Lottermoser}},
  \bibinfo{author}{\bibfnamefont{T.}~\bibnamefont{Lonkai}},
  \bibinfo{author}{\bibfnamefont{U.}~\bibnamefont{Amann}},
  \bibinfo{author}{\bibfnamefont{D.}~\bibnamefont{Hohlwein}},
  \bibinfo{author}{\bibfnamefont{J.}~\bibnamefont{Ihringer}}, \bibnamefont{and}
  \bibinfo{author}{\bibfnamefont{M.}~\bibnamefont{Fiebig}},
  \bibinfo{journal}{Nature} \textbf{\bibinfo{volume}{430}},
  \bibinfo{pages}{541} (\bibinfo{year}{2004}).

\bibitem[{\citenamefont{Fiebig et~al.}(2001)\citenamefont{Fiebig, Fr{\"o}hlich,
  Lottermoser, Pavlov, Pisarev, and Weber}}]{manfred}
\bibinfo{author}{\bibfnamefont{M.}~\bibnamefont{Fiebig}},
  \bibinfo{author}{\bibfnamefont{D.}~\bibnamefont{Fr{\"o}hlich}},
  \bibinfo{author}{\bibfnamefont{T.}~\bibnamefont{Lottermoser}},
  \bibinfo{author}{\bibfnamefont{V.~V.} \bibnamefont{Pavlov}},
  \bibinfo{author}{\bibfnamefont{R.~V.} \bibnamefont{Pisarev}},
  \bibnamefont{and} \bibinfo{author}{\bibfnamefont{H.-J.} \bibnamefont{Weber}},
  \bibinfo{journal}{Phys. Rev. Lett.} \textbf{\bibinfo{volume}{87}},
  \bibinfo{pages}{137202} (\bibinfo{year}{2001}).

\bibitem[{\citenamefont{Lefkidis and H{\"u}bner}(2005)}]{lefkidis}
\bibinfo{author}{\bibfnamefont{G.}~\bibnamefont{Lefkidis}} \bibnamefont{and}
  \bibinfo{author}{\bibfnamefont{W.}~\bibnamefont{H{\"u}bner}},
  \bibinfo{journal}{Phys. Rev. Lett.} \textbf{\bibinfo{volume}{95}},
  \bibinfo{pages}{77401} (\bibinfo{year}{2005}).

\bibitem[{\citenamefont{Satitkovitchai
  et~al.}(2005)\citenamefont{Satitkovitchai, Pavlyukh, and
  H{\"u}bner}}]{KS_SOC}
\bibinfo{author}{\bibfnamefont{K.}~\bibnamefont{Satitkovitchai}},
  \bibinfo{author}{\bibfnamefont{Y.}~\bibnamefont{Pavlyukh}}, \bibnamefont{and}
  \bibinfo{author}{\bibfnamefont{W.}~\bibnamefont{H{\"u}bner}},
  \bibinfo{journal}{Phys. Rev. B} \textbf{\bibinfo{volume}{72}},
  \bibinfo{pages}{45116} (\bibinfo{year}{2005}).

\bibitem[{\citenamefont{N{\'y}vlt et~al.}(2008)\citenamefont{N{\'y}vlt, Bisio,
  and Kirschner}}]{SHG2008}
\bibinfo{author}{\bibfnamefont{M.}~\bibnamefont{N{\'y}vlt}},
  \bibinfo{author}{\bibfnamefont{F.}~\bibnamefont{Bisio}}, \bibnamefont{and}
  \bibinfo{author}{\bibfnamefont{J.}~\bibnamefont{Kirschner}},
  \bibinfo{journal}{Phys. Rev. B} \textbf{\bibinfo{volume}{77}},
  \bibinfo{pages}{14435} (\bibinfo{year}{2008}).

\bibitem[{\citenamefont{Fromme et~al.}(1996)\citenamefont{Fromme, M{\"o}ller,
  Ansch{\"u}tz, Bethke, and Kisker}}]{fromme}
\bibinfo{author}{\bibfnamefont{B.}~\bibnamefont{Fromme}},
  \bibinfo{author}{\bibfnamefont{M.}~\bibnamefont{M{\"o}ller}},
  \bibinfo{author}{\bibfnamefont{T.}~\bibnamefont{Ansch{\"u}tz}},
  \bibinfo{author}{\bibfnamefont{C.}~\bibnamefont{Bethke}}, \bibnamefont{and}
  \bibinfo{author}{\bibfnamefont{E.}~\bibnamefont{Kisker}},
  \bibinfo{journal}{Phys. Rev. Lett.} \textbf{\bibinfo{volume}{77}},
  \bibinfo{pages}{1548} (\bibinfo{year}{1996}).

\bibitem[{\citenamefont{Gorschl{\"u}ter and Merz}(1994)}]{001_EELS}
\bibinfo{author}{\bibfnamefont{A.}~\bibnamefont{Gorschl{\"u}ter}}
  \bibnamefont{and} \bibinfo{author}{\bibfnamefont{H.}~\bibnamefont{Merz}},
  \bibinfo{journal}{Phys. Rev. B} \textbf{\bibinfo{volume}{49}},
  \bibinfo{pages}{17293} (\bibinfo{year}{1994}).

\bibitem[{\citenamefont{Newman and Chrenko}(1959)}]{optical_bulk}
\bibinfo{author}{\bibfnamefont{R.}~\bibnamefont{Newman}} \bibnamefont{and}
  \bibinfo{author}{\bibfnamefont{R.}~\bibnamefont{Chrenko}},
  \bibinfo{journal}{Phys. Rev.} \textbf{\bibinfo{volume}{114}},
  \bibinfo{pages}{1507} (\bibinfo{year}{1959}).

\bibitem[{\citenamefont{Lichtenstein and Katsnelson}(1998)}]{katsnelson}
\bibinfo{author}{\bibfnamefont{A.~I.} \bibnamefont{Lichtenstein}}
  \bibnamefont{and} \bibinfo{author}{\bibfnamefont{M.~I.}
  \bibnamefont{Katsnelson}}, \bibinfo{journal}{Phys. Rev. B}
  \textbf{\bibinfo{volume}{57}}, \bibinfo{pages}{6884} (\bibinfo{year}{1998}).

\bibitem[{\citenamefont{Aryasetiawan and Gunnarsson}(1995)}]{aryasetiawan}
\bibinfo{author}{\bibfnamefont{F.}~\bibnamefont{Aryasetiawan}}
  \bibnamefont{and}
  \bibinfo{author}{\bibfnamefont{O.}~\bibnamefont{Gunnarsson}},
  \bibinfo{journal}{Phys. Rev. Lett.} \textbf{\bibinfo{volume}{74}},
  \bibinfo{pages}{3221} (\bibinfo{year}{1995}).

\bibitem[{\citenamefont{Faleev et~al.}(2004)\citenamefont{Faleev,
  v.~Schilfgaarde, and Kotani}}]{faleev}
\bibinfo{author}{\bibfnamefont{S.~V.} \bibnamefont{Faleev}},
  \bibinfo{author}{\bibfnamefont{M.}~\bibnamefont{v.~Schilfgaarde}},
  \bibnamefont{and} \bibinfo{author}{\bibfnamefont{T.}~\bibnamefont{Kotani}},
  \bibinfo{journal}{Phys. Rev. Lett.} \textbf{\bibinfo{volume}{93}},
  \bibinfo{pages}{126406} (\bibinfo{year}{2004}).

\bibitem[{\citenamefont{Li et~al.}(2005)\citenamefont{Li, Rignanese, and
  Louie}}]{louie}
\bibinfo{author}{\bibfnamefont{J.-L.} \bibnamefont{Li}},
  \bibinfo{author}{\bibfnamefont{G.-M.} \bibnamefont{Rignanese}},
  \bibnamefont{and} \bibinfo{author}{\bibfnamefont{S.}~\bibnamefont{Louie}},
  \bibinfo{journal}{Phys. Rev. B} \textbf{\bibinfo{volume}{71}},
  \bibinfo{pages}{193102} (\bibinfo{year}{2005}).

\bibitem[{\citenamefont{Kune{\v{s}} et~al.}(2007)\citenamefont{Kune{\v{s}},
  Anisimov, Skornyakov, Lukoyanov, and Vollhardt}}]{VollhardtPRL}
\bibinfo{author}{\bibfnamefont{J.}~\bibnamefont{Kune{\v{s}}}},
  \bibinfo{author}{\bibfnamefont{V.~I.} \bibnamefont{Anisimov}},
  \bibinfo{author}{\bibfnamefont{S.~L.} \bibnamefont{Skornyakov}},
  \bibinfo{author}{\bibfnamefont{A.~V.} \bibnamefont{Lukoyanov}},
  \bibnamefont{and}
  \bibinfo{author}{\bibfnamefont{D.}~\bibnamefont{Vollhardt}},
  \bibinfo{journal}{Phys. Rev. Lett.} \textbf{\bibinfo{volume}{99}},
  \bibinfo{pages}{156404} (\bibinfo{year}{2007}).

\bibitem[{\citenamefont{Satitkovitchai
  et~al.}(2003)\citenamefont{Satitkovitchai, Pavlyukh, and
  H{\"u}bner}}]{khompat}
\bibinfo{author}{\bibfnamefont{K.}~\bibnamefont{Satitkovitchai}},
  \bibinfo{author}{\bibfnamefont{Y.}~\bibnamefont{Pavlyukh}}, \bibnamefont{and}
  \bibinfo{author}{\bibfnamefont{W.}~\bibnamefont{H{\"u}bner}},
  \bibinfo{journal}{Phys. Rev. B} \textbf{\bibinfo{volume}{67}},
  \bibinfo{pages}{165413} (\bibinfo{year}{2003}).

\bibitem[{\citenamefont{Head-Gordon et~al.}(1995)\citenamefont{Head-Gordon,
  Maurice, and Oumi}}]{cisd2}
\bibinfo{author}{\bibfnamefont{M.}~\bibnamefont{Head-Gordon}},
  \bibinfo{author}{\bibfnamefont{D.}~\bibnamefont{Maurice}}, \bibnamefont{and}
  \bibinfo{author}{\bibfnamefont{M.}~\bibnamefont{Oumi}},
  \bibinfo{journal}{Chem. Phys. Lett.} \textbf{\bibinfo{volume}{246}},
  \bibinfo{pages}{114} (\bibinfo{year}{1995}).

\bibitem[{\citenamefont{Lefkidis and H{\"u}bner}(2006)}]{lefkidisPRB}
\bibinfo{author}{\bibfnamefont{G.}~\bibnamefont{Lefkidis}} \bibnamefont{and}
  \bibinfo{author}{\bibfnamefont{W.}~\bibnamefont{H{\"u}bner}},
  \bibinfo{journal}{Phys. Rev. B} \textbf{\bibinfo{volume}{74}},
  \bibinfo{pages}{155106} (\bibinfo{year}{2006}).

\bibitem[{\citenamefont{from \cite{lefkidis} and the~references
  therein.}()}]{text}
\bibinfo{author}{\bibfnamefont{T.}~\bibnamefont{from \cite{lefkidis}}}
  \bibnamefont{and} \bibinfo{author}{\bibnamefont{the~references therein.}}
  (????).

\bibitem[{\citenamefont{Tanabe et~al.}(2000)\citenamefont{Tanabe, Fiebig, and
  Hanamura}}]{sugano}
\bibinfo{author}{\bibfnamefont{Y.}~\bibnamefont{Tanabe}},
  \bibinfo{author}{\bibfnamefont{M.}~\bibnamefont{Fiebig}}, \bibnamefont{and}
  \bibinfo{author}{\bibfnamefont{E.}~\bibnamefont{Hanamura}},
  \emph{\bibinfo{title}{Magnetoptics}} (\bibinfo{publisher}{Springer},
  \bibinfo{year}{2000}).

\bibitem[{\citenamefont{Shen}(1984)}]{shen}
\bibinfo{author}{\bibfnamefont{Y.~R.} \bibnamefont{Shen}},
  \emph{\bibinfo{title}{The Principles of Nonlinear Optics}}
  (\bibinfo{publisher}{John Wiley and Sons}, \bibinfo{year}{1984}).

\bibitem[{\citenamefont{Andersen et~al.}(2004)\citenamefont{Andersen, Keller,
  H\"ubner, and Johansson}}]{TA-nonlinear}
\bibinfo{author}{\bibfnamefont{T.}~\bibnamefont{Andersen}},
  \bibinfo{author}{\bibfnamefont{O.}~\bibnamefont{Keller}},
  \bibinfo{author}{\bibfnamefont{W.}~\bibnamefont{H\"ubner}}, \bibnamefont{and}
  \bibinfo{author}{\bibfnamefont{B.}~\bibnamefont{Johansson}},
  \bibinfo{journal}{Phys. Rev. A} \textbf{\bibinfo{volume}{70}},
  \bibinfo{pages}{043806} (\bibinfo{year}{2004}).

\bibitem[{\citenamefont{H{\"u}bner and Bennemann}(1989)}]{tensorformula}
\bibinfo{author}{\bibfnamefont{W.}~\bibnamefont{H{\"u}bner}} \bibnamefont{and}
  \bibinfo{author}{\bibfnamefont{K.-H.} \bibnamefont{Bennemann}},
  \bibinfo{journal}{Phys. Rev. B} \textbf{\bibinfo{volume}{40}},
  \bibinfo{pages}{5973} (\bibinfo{year}{1989}).

\bibitem[{\citenamefont{Lefkidis et~al.}(2005)\citenamefont{Lefkidis, Ney, and
  H{\"u}bner}}]{aalborg}
\bibinfo{author}{\bibfnamefont{G.}~\bibnamefont{Lefkidis}},
  \bibinfo{author}{\bibfnamefont{O.}~\bibnamefont{Ney}}, \bibnamefont{and}
  \bibinfo{author}{\bibfnamefont{W.}~\bibnamefont{H{\"u}bner}},
  \bibinfo{journal}{Phys. Stat. Sol. (C)} \textbf{\bibinfo{volume}{2}},
  \bibinfo{pages}{4022} (\bibinfo{year}{2005}).

\bibitem[{\citenamefont{D{\"a}hn et~al.}(1996)\citenamefont{D{\"a}hn,
  H{\"u}bner, and Bennemann}}]{daehnPRL}
\bibinfo{author}{\bibfnamefont{A.}~\bibnamefont{D{\"a}hn}},
  \bibinfo{author}{\bibfnamefont{W.}~\bibnamefont{H{\"u}bner}},
  \bibnamefont{and} \bibinfo{author}{\bibfnamefont{K.~H.}
  \bibnamefont{Bennemann}}, \bibinfo{journal}{Phys. Rev. Lett.}
  \textbf{\bibinfo{volume}{77}}, \bibinfo{pages}{3929} (\bibinfo{year}{1996}).

\bibitem[{\citenamefont{Birss}(1964)}]{birss}
\bibinfo{author}{\bibfnamefont{R.}~\bibnamefont{Birss}},
  \emph{\bibinfo{title}{Symmetry and Magnetism}}
  (\bibinfo{publisher}{North-Holland Publishing Company},
  \bibinfo{year}{1964}).

\bibitem[{\citenamefont{Andersen and H{\"u}bner}(2002)}]{AndersenSHG}
\bibinfo{author}{\bibfnamefont{T.}~\bibnamefont{Andersen}} \bibnamefont{and}
  \bibinfo{author}{\bibfnamefont{W.}~\bibnamefont{H{\"u}bner}},
  \bibinfo{journal}{Phys. Rev. B} \textbf{\bibinfo{volume}{65}},
  \bibinfo{pages}{174409} (\bibinfo{year}{2002}).

\bibitem[{\citenamefont{G{\'o}mez-Abal and H{\"u}bner}(2002)}]{abalPRB}
\bibinfo{author}{\bibfnamefont{R.}~\bibnamefont{G{\'o}mez-Abal}}
  \bibnamefont{and}
  \bibinfo{author}{\bibfnamefont{W.}~\bibnamefont{H{\"u}bner}},
  \bibinfo{journal}{Phys. Rev. B} \textbf{\bibinfo{volume}{65}},
  \bibinfo{pages}{195114} (\bibinfo{year}{2002}).

\bibitem[{\citenamefont{G{\'o}mez-Abal and H{\"u}bner}(2003)}]{abalJPCM}
\bibinfo{author}{\bibfnamefont{R.}~\bibnamefont{G{\'o}mez-Abal}}
  \bibnamefont{and}
  \bibinfo{author}{\bibfnamefont{W.}~\bibnamefont{H{\"u}bner}},
  \bibinfo{journal}{J. Phys. Condens. Matter} \textbf{\bibinfo{volume}{15}},
  \bibinfo{pages}{S709} (\bibinfo{year}{2003}).

\bibitem[{\citenamefont{Longa et~al.}(2006)\citenamefont{Longa, Kohlhepp,
  de~Jonge, and Koopmans}}]{koopmanscondmat}
\bibinfo{author}{\bibfnamefont{F.~D.} \bibnamefont{Longa}},
  \bibinfo{author}{\bibfnamefont{C.~J.~T.} \bibnamefont{Kohlhepp}},
  \bibinfo{author}{\bibfnamefont{W.~J.~M.} \bibnamefont{de~Jonge}},
  \bibnamefont{and} \bibinfo{author}{\bibfnamefont{B.}~\bibnamefont{Koopmans}},
  \bibinfo{journal}{cond-mat} p. \bibinfo{pages}{0609698}
  (\bibinfo{year}{2006}).

\bibitem[{\citenamefont{Lefkidis and H{\"u}bner}(2007)}]{lefkidisswitch}
\bibinfo{author}{\bibfnamefont{G.}~\bibnamefont{Lefkidis}} \bibnamefont{and}
  \bibinfo{author}{\bibfnamefont{W.}~\bibnamefont{H{\"u}bner}},
  \bibinfo{journal}{Phys. Rev. B} \textbf{\bibinfo{volume}{76}},
  \bibinfo{pages}{014418} (\bibinfo{year}{2007}).

\bibitem[{\citenamefont{Stanciu et~al.}(2007)\citenamefont{Stanciu, Hansteen,
  Kimel, Kirilyuk, Tsukamoto, Itoh, and Rasing}}]{KimelPRL}
\bibinfo{author}{\bibfnamefont{C.~D.} \bibnamefont{Stanciu}},
  \bibinfo{author}{\bibfnamefont{F.}~\bibnamefont{Hansteen}},
  \bibinfo{author}{\bibfnamefont{A.~V.} \bibnamefont{Kimel}},
  \bibinfo{author}{\bibfnamefont{A.}~\bibnamefont{Kirilyuk}},
  \bibinfo{author}{\bibfnamefont{A.}~\bibnamefont{Tsukamoto}},
  \bibinfo{author}{\bibfnamefont{A.}~\bibnamefont{Itoh}}, \bibnamefont{and}
  \bibinfo{author}{\bibfnamefont{T.}~\bibnamefont{Rasing}},
  \bibinfo{journal}{Phys. Rev. Lett.} \textbf{\bibinfo{volume}{99}},
  \bibinfo{pages}{047601} (\bibinfo{year}{2007}).

\bibitem[{\citenamefont{Lefkidis and H{\"u}bner}(in press)}]{QuantizedPhonon}
\bibinfo{author}{\bibfnamefont{G.}~\bibnamefont{Lefkidis}} \bibnamefont{and}
  \bibinfo{author}{\bibfnamefont{W.}~\bibnamefont{H{\"u}bner}},
  \bibinfo{journal}{J. Magn. Magn. Mater.}  (\bibinfo{year}{in press}).

\end{thebibliography}

\end{document}